\documentclass[conference]{IEEEtran}

\usepackage{color}
\usepackage{cite}
\usepackage{soul}
\usepackage{balance}

\ifCLASSINFOpdf
\else
  \usepackage[dvips]{graphicx}
  \graphicspath{{/}}
\fi
\usepackage{epstopdf}

\usepackage{nicefrac}
\usepackage{amsmath, amssymb, amsthm}
\usepackage{mathtools}

\usepackage{algorithm}
\usepackage{algorithmic}

\usepackage{booktabs}
\usepackage[caption=false]{subfig}

\usepackage{url}
\newcommand*{\QEDA}{\hfill\ensuremath{\blacksquare}}%

\begin{document}

\title{Cross-Layer Optimization of Wireless Links \\ under Reliability and Energy Constraints}

\author{\IEEEauthorblockN{Aamir~Mahmood\IEEEauthorrefmark{1},
M~M~Aftab~Hossain\IEEEauthorrefmark{2} and Mikael~Gidlund\IEEEauthorrefmark{1}}
\IEEEauthorblockA{\IEEEauthorrefmark{1} Department of Information Systems and Technology, Mid Sweden University, Sweden}
\IEEEauthorblockA{\IEEEauthorrefmark{2} Wireless@KTH, KTH Royal Institute of Technology, Sweden}
Email: \IEEEauthorrefmark{1}firstname.lastname@miun.se, \IEEEauthorrefmark{2}mmaho@kth.se
}
\maketitle

\begin{abstract}
   

The vision of connecting billions of battery operated devices to be used for 
diverse emerging applications calls for a wireless communication system that 
can support stringent reliability and latency requirements. Both reliability 
and energy efficiency are critical for many of these applications that 
involve communication with short packets which undermine the coding gain 
achievable from large packets. 
In this paper, we study a cross-layer approach to optimize the performance of low-power wireless links.
At first, we derive a simple and accurate packet error rate (PER) expression for 
uncoded schemes in block fading channels based on a new proposition that 
shows that the waterfall threshold in the PER upper bound in Nakagami-$m$ 
fading channels is tightly approximated by the $m$-th moment of an 
asymptotic distribution of PER in AWGN channel. The proposed PER approximation establishes an 
explicit connection between the physical and link 
layers parameters, and the packet error rate. We exploit this connection for cross-layer design and 
optimization of communication links. To this end, we propose a semi-analytic framework to jointly optimize signal-to-noise 
ratio (SNR) and modulation order at physical layer, and the packet length and 
number of retransmissions at link layer with respect to distance under the 
prescribed delay and reliability constraints. 
\end{abstract}

\IEEEpeerreviewmaketitle


\section{Introduction} 

The upcoming wireless networks are required to support massive number of 
devices under the umbrella of internet-of-things (IoT).  The heterogeneity of 
use cases of these  machine-to-machine (M2M) type communication necessitate 
diverse reliability and latency requirements. As the devices will be 
mostly battery operated, energy efficiency also becomes a critical issue.  
Moreover, this novel traffic type uses short packets which undermine the 
coding gain achievable from large packets \cite{ShortPackets}\cite{Shannon}. All these factors urge a 
new look not only into physical layer but also cross-layer design in order to 
ensure reliability under energy constraints. In this paper, we improve the 
packet error rate (PER) approximations over block-fading channels in order to 
have a better control over the parameters that determine the system 
performance and utilize these insights to optimize cross-layer 
parameters, e.g., packet length, number of retransmission, modulation scheme.

The average PER is an important metric for cross-layer 
optimization of wireless transmissions over block-fading channels. 
For instance, the objective function to optimize throughput, energy or 
spectral efficiency of a transmission scheme is defined in relation to the 
average PER \cite{liu2004cross}\cite{rosas2012modulation}, and the parameters 
maximizing the system performance are determined. However, the average PER, except for certain simple cases, is not found in exact 
closed form, although it can usually be written in the integral form. The 
integral then needs to be evaluated numerically and may not be 
computationally intensive, however this approach in general does not offer 
insights as to what parameters determine the system performance.


One such closed form is the upper bound on average PER for both the uncoded 
and coded schemes in Rayleigh fading, $1-\exp(\omega_0/\bar{\gamma})$, where $
\bar{\gamma}$ is the average signal-to-noise 
ratio (SNR) and $\omega_0$ is the waterfall threshold \cite{xi2011general}. The threshold is 
defined as an integral of the PER function in the AWGN channel. In 
\cite{xi2011general}, also a similar upper bound in Nakagami-$m$ fading is 
proposed with a corresponding threshold. In both cases, however, a 
closed-form solution to the threshold is not feasible. A log-domain linear 
approximation of $\omega_0$ is developed for uncoded schemes in 
\cite{liu2012tput}, and for (un)coded schemes in \cite{wu2014energy}. 
However, the approximation in \cite{liu2012tput} is tight for large packets 
only while in \cite{wu2014energy} the approximation parameters for a given modulation scheme are
calculated by simulations. For uncoded schemes,
an accurate PER expression is derived in \cite{Our2016} that is
complicated to utilize for link optimization. 

In this paper, we show that the waterfall threshold in the PER upper bound under 
Nakagami-$m$ block fading is tightly approximated by the $m$-th moment of an 
asymptotic distribution of PER function in AWGN. In Rayleigh fading, the approximation leads to 
a PER approximation which is accurate than 
\cite{liu2012tput}\cite{wu2014energy} and also maintains explicit connection with 
modulation order unlike \cite{wu2014energy}.


Note that in today's age of battery operated devices, the number one design 
goal is energy efficient communication. However, the emerging delay and 
reliability requirements have a direct impact on the needed energy to transfer 
each information bit. The reliability depends on the bit error or packet 
error statistics of the wireless channel, which in turn depend on the choice 
of the system parameters such as transmit power, modulation scheme, packet 
length etc. If the packet error probability has to be reduced so as to 
transfer a packet with limited number of retransmission, these 
parameters need to be optimized while keeping a tab on the energy 
consumption.  

How to select the modulation order and transmission power to attain 
energy-efficient communication is studied in AWGN channel 
\cite{cui2005energy, wang2008, hou2005performance} and in 
fading channels~\cite{rosas2012modulation}\cite{wu2014energy}. 
These studies in general suggest using higher order modulations at 
smaller distances as opposed to the common notion followed in wireless sensor 
networks (WSN) by choosing 
low-order modulations for their low SNR requirement. For instance, low-power 
transceivers CC1100 and CC2420, often 
used in WSNs, employ BPSK and QPSK. In fading channels, it is shown in 
\cite{rosas2012modulation}\cite{wu2014energy} that there 
exist an optimal SNR and packet length for each modulation scheme at which 
the required energy for successful transfer of an information bit is 
minimized. In \cite{rosas2012modulation}, the optimal SNR is conditioned on 
the maximum transmit power however this constraint is ignored in 
\cite{wu2014energy}. In these studies, no restriction on 
number of retransmission is imposed and as a result the optimal SNR is not bound to satisfy the reliability target. 
%

In this paper, we study the energy minimization in fading channels however 
under the often neglected reliability constraints. We exploit the proposed 
PER approximation for cross-layer optimization of a power-limited system in 
Rayleigh block-fading channels. By defining a energy consumption model for 
per payload bit transferred, we find the optimal (energy consumption 
minimizing) system parameters while maintaining the reliability and delay 
target. Specifically, i) for a system with fixed modulation scheme (e.g., 
CC2420) and report size, we propose closed-form conditions for energy optimal 
SNR that conform to the maximum transmit power and reliability constraints, 
ii) for a general power-limited system, we propose a joint optimization 
algorithm to find the physical layer (SNR, modulation order) and link layer 
(packet length, number of retransmissions) parameters with respect to 
distance under the prescribed delay and reliability constraints.
%

%

The rest of the paper is organized as follows. Section \ref{sec:PER} develops 
an approximation to average PER in block fading channels. Section \ref{sec:EE} 
defines the cross-layer optimization problem, solves it under the 
reliability constraints and presents the results. Section 
\ref{sec:Conclusions} draws the concluding remarks.

\section{The Average PER in Block Fading}
\label{sec:PER}
Let $f\left(\gamma\right)$ be the PER function in the AWGN channel with 
instantaneous SNR, $\gamma$. Then, for an $N$-bit uncoded packet with bit error rate (BER) 
function $b_e\left(\gamma\right)$, $f\left(\gamma\right)$ is defined as
\begin{equation}
\vspace{-2pt}
f(\gamma) = 1-\Big(1-b_e\left(\gamma\right)\Big)^{N}
\label{eq:p_awgn1}
\end{equation} 
Also, let $p\left(\gamma;\bar{\gamma}\right)$ be the probability distribution 
function (PDF) of the received instantaneous SNR. In Nakagami-$m$ fading, 
$\gamma$ follows the Gamma distribution with PDF
\begin{equation}
p(\gamma;\bar{\gamma}) = \frac{m^m \gamma^{m-1}}{\bar{\gamma}^m\Gamma(m)}\mathrm{exp}\Big(-
\frac{m\gamma}{\bar{\gamma}}\Big), \gamma \geq 0
\label{eq:Gamma_Distr}
\end{equation}   
The average PER, denoted as $\bar{P_e}\left(\bar{\gamma}\right)$, is then computed 
by integrating \eqref{eq:p_awgn1} over \eqref{eq:Gamma_Distr}
\begin{equation}
\bar{P_e}\left(\gamma\right) = \int_0^{\infty}f(\gamma){p(\gamma;\bar{\gamma})} d\gamma
\label{eq:ref_PER}
\end{equation}   

In \cite{xi2011general}, it is shown that $\bar{P_e}\left(\bar{\gamma}\right)$ is upper bounded by
\begin{equation}
\bar{P_e}\left(\gamma\right) \leq \frac{m^{m-1}B}{\bar{\gamma}^{m-1}\Gamma\left(m\right)}
\left(1-\mathrm{exp}\left(-\frac{m\omega_m}{\bar{\gamma}B}\right)\right) 
\label{eq:UB_Naka}
\end{equation}
where $0\leq \gamma^{m-1}f\left(\gamma\right)\leq B$ and $\omega_m$ is defined as,
\begin{equation}
\omega_m = \int_0^\infty {\gamma^{m-1}f\left(\gamma\right)d\gamma} 
\label{eq:threshold}
\end{equation}   

In Rayleigh fading (i.e., $m=1$), as $f\left(\gamma\right)$ is the probability we have 
$0\leq f\left(\gamma\right) \leq 1$, and \eqref{eq:UB_Naka} can be written as,
\begin{equation}
\bar{P_e}\left(\gamma\right) \leq 1-\mathrm{exp}\left(-\frac{\omega_0}{\bar{
\gamma}}\right) 
\label{eq:UB_Ray}
\end{equation}
where $\omega_0$ from \eqref{eq:threshold} becomes
\begin{equation}
\omega_0 = \int_0^\infty {f\left(\gamma\right)d\gamma} 
\label{eq:threshold_0}
\end{equation}   

In what follows, we propose generic approximations to $\omega_0$
 and $\omega_m$ for uncoded schemes with BER functions as
\begin{equation} 	
b_{e}(\gamma) = c_m\mathrm{exp}\left(-k_m\gamma\right) 
\label{eq:f_se}
\end{equation} 
\begin{equation} 	
b_{e}(\gamma) = c_mQ\left(\sqrt{k_m\gamma}\right) 
\label{eq:f_sq}
\end{equation} 
where $c_m$ and $k_m$ are modulation-dependent constants. Non-coherent FSK 
and DPSK have the BER in the form of \eqref{eq:f_se} while M-ASK, M-PAM, 
MSK, M-PSK and M-QAM have BER in the Gaussian $Q$-function form \eqref{eq:f_sq}\cite{xi2011general}.

\subsection{Approximations to $\omega_0$ and $\omega_m$}
\textit{Proposition 1}: For uncoded transmission of a packet with length $N$, 
with the BER functions described by $c_m e^{-k_m\gamma}$ and 
$c_m Q(\sqrt{k_m\gamma})$ where $0 < c_m \leq 1$ and $k_m >0$, the 
threshold, $\omega_m$, in Nakagami-$m$ fading channel for integer values of the fading 
parameter is approximated 
by the \textit{m}th moment of the Gumbel distribution for sample maximum 
\begin{equation}
\omega_m \approx \frac{\mathbb{E}[\gamma^m]}{m}
\label{eq:proposedThreshold}
\end{equation}

\textit{Proof}: For packet length $N$, the PER function in 
\eqref{eq:p_awgn1} for BER functions described by $c_m e^{-k_m\gamma}$ and 
$c_m Q(\sqrt{k_m\gamma})$ can be asymptotically approximated by the Gumbel 
distribution function for the sample minimum \cite{Our2016}
\begin{equation}
f(\gamma) \simeq 1-\mathrm{exp}\left(-\mathrm{exp}\left(-\frac{\gamma-a_N}{b_N
}\right)\right)
\label{eq:gumbelDistributionApprox}
\end{equation}
where $a_N$ and $b_N>0$ are the normalizing constants.

Let $G\left(\gamma\right) = \mathrm{exp}(-\mathrm{exp}(-
\frac{\gamma-a_N}{b_N}))$ be the cumulative distribution function (CDF) of 
the Gumbel distribution for the sample maximum, then from 
\eqref{eq:gumbelDistributionApprox} and \eqref{eq:threshold} we have
\begin{equation}
\omega_m \approx \int_0^\infty {\gamma^{m-1}\Big(1-G\left(\gamma\right)\Big)d
\gamma} 
\label{eq:threshold_new}
\end{equation} 
Assuming $\gamma = y^{\frac{1}{m}}$ and $\gamma^{m-1}d\gamma = {dy}/{m}$, from \eqref{eq:threshold_new} we get
\begin{equation}
\omega_m \approx \frac{1}{m} \int_0^\infty {1-G\left(y^{\frac{1}{m}}\right)dy} 
\label{eq:threshold_new_temp1}
\end{equation} 

Let $g(\gamma) = {dG(\gamma)}/{d\gamma}$ be the PDF, then with some manipulation and changing the 
order of integration in \eqref{eq:threshold_new_temp1}
\begin{align}
\omega_m & \approx \frac{1}{m}\int_0^\infty{\int_{y^{\frac{1}{m}}}^{\infty}g
\left(\gamma\right)d\gamma dy} \nonumber \\
				 &  = \frac{1}{m}\int_0^\infty{\int_0^{\gamma^m}g\left(\gamma\right)
dy d\gamma} \nonumber \\ 
				 &  = \frac{1}{m}\int_0^\infty{\gamma^m g(\gamma)d\gamma}.	
\label{eq:lemma1_2}
\end{align}
Noting that the integral in the last equality is the $m$th moment of a 
continuous and nonnegative random variable $\gamma$ with the PDF 
$g(\gamma)$ completes the proof.\QEDA

One can find the \textit{m}th moment of the Gumbel distribution from its 
moment generating function (MGF) defined as
\begin{equation}
M_{\gamma}\left(t\right) \triangleq \Gamma \Big(1-b_N \,t\Big)\,e^{{a_N \,t}}  
\label{eq:MGF_Gumbel}
\end{equation} 
where $\Gamma\left(\cdot\right)$ is the standard gamma function.
 
In Rayleigh fading with $m=1$, from \eqref{eq:proposedThreshold} and 
\eqref{eq:MGF_Gumbel}, $\omega_0$ equals the expected value of the Gumbel distribution, 
i.e.
\begin{equation}
\omega_0 \approx \mathbb{E}[\gamma] = a_N +  \gamma_e \, b_N
\label{eq:solution_w0}
\end{equation}
where $\gamma_e = 0.5772$ is the Euler constant. Notation $\omega_0$ is 
preferred over $\omega_1$ to remain consistent with the prior works. Similarly, for $m=2$ and 
$m=3$, which represent the next two significant fading conditions, 
\eqref{eq:proposedThreshold} under \eqref{eq:MGF_Gumbel} becomes
\begin{align}
\begin{split}\label{eq:1}
		\omega_2 \! \approx \! \frac{\mathbb{E}[\gamma^2]}{2} \! = & \frac{1}{2}\Big[a_N^2 + 1.64b_{N}^2 + \gamma_e^2 b_{N}^2 + 2\gamma_e a_N b_N\Big]
\end{split}\\
\begin{split}\label{eq:2}
		\omega_3 \! \approx \! \frac{\mathbb{E}[\gamma^3]}{3} \! = & \frac{1}{3} \Big[4.93\gamma_e b_{N}^3 \!+\! 4.93 a_{N} b_N^2 \!+\! a_{N}^3 \!+\! 2.40 b_{N}^3
																										\\ & \quad \,\,\, + \gamma_e^3b_{N}^3+ 3\gamma_e^2a_N b_N^2 + 3\gamma_e a_N^2 b_N \Big]
\end{split}
\end{align}
The normalizing constants for BER function in \eqref{eq:f_se} are \cite{Our2016}
\begin{equation}
a_N = \frac{\log (Nc_m)}{k_m},\,\,\,\,\,\,\,\, b_N = \frac{1}{k_m}
\label{eq:AnBn_expo}
\end{equation}
whereas the constants for BER in \eqref{eq:f_sq} are
\begin{equation}
\begin{aligned}
a_N & = \frac{2}{k_m} \Big[\mathrm{erf}^{-1}\Big(1 - \frac{2}{Nc_m}\Big)\Big]^
2  \\
b_N & = \frac{2}{k_m} \Big[\mathrm{erf}^{-1}\Big(1 - \frac{2}{Nc_me}\Big)\Big]
^2 - a_N 
\label{eq:AnBn_Qfunc}
\end{aligned}
\end{equation}
where $e$ is the base of the natural logarithm. 

\subsection{The Average PER with New Parametrization}

Using the normalizing constants \eqref{eq:AnBn_expo} in 
\eqref{eq:proposedThreshold}, the average PER in \eqref{eq:UB_Naka} and 
\eqref{eq:UB_Ray} can be expressed in the form of elementary functions. 
However due to the inverse error function in 
\eqref{eq:AnBn_Qfunc}, 
\eqref{eq:UB_Naka} and \eqref{eq:UB_Ray} cannot be simplified 
further. An intuitive approach is to utilize an exponential function based 
approximation of $Q$-function (e.g., \cite{wu2011new}), and 
utilize $a_N$ and $b_N$ from \eqref{eq:AnBn_expo}. However, this approach 
loses the approximation accuracy. Instead, our  objective is to find the 
exponential function approximation for given $a_N$ and $b_N$ in \eqref{eq:AnBn_expo} and $\omega_m$ approximation in 
\eqref{eq:proposedThreshold} that fits best to the integral expression in 
\eqref{eq:threshold} or \eqref{eq:threshold_0}. In essence, we reformulate 
$a_N$ and $b_N$ in \eqref{eq:AnBn_expo} as    
\begin{equation}
a_N \approx \frac{\log (k_1Nc_m)}{k_2k_m},\,\,\,\,\,\,\,\, b_N \approx \frac{1
}{k_2k_m}
\label{eq:AnBn_expo2}
\end{equation}  
and find the constants $k_1$ and $k_2$. We estimated $k_1$ and $k_2$ for BPSK 
modulation by numerically evaluating \eqref{eq:threshold_0} and matching it
 with \eqref{eq:solution_w0} under $a_N$ and $b_N$ 
in \eqref{eq:AnBn_expo2}. For a packet length $N$ in an interval $[32, 1024]$ bits, the 
optimal constants are: $k_1= 0.2114$ and 
$k_2 = 0.5598$. We find that these constants are independent of modulation 
schemes with the BER function involving $Q$-function. As a 
result, a simple PER approximation can also be reached for the 
modulation schemes with the BER function $c_m Q(\sqrt{k_m\gamma})$. For 
instance, let $\acute{c}_m = k_1 c_m$ and $\acute{k}_m = k_2 k_m$, 
then from \eqref{eq:AnBn_expo2} and \eqref{eq:solution_w0}, the PER 
in Rayleigh fading is
\begin{equation}
\bar{P}_{e}(\bar{\gamma}) \approx 1-\left(N\acute{c}_m\right)^{-\frac{1}{
\acute{k}_m\bar{\gamma}}} \mathrm{exp}\left(-\frac{\gamma_e}{\acute{k}_m\bar{
\gamma}}\right).
\label{eq:ubPER1}
\end{equation}
where $\acute{c}_m = c_m$ and $\acute{k}_m = k_m$ for the BER function $c_m e^
{-k_m\gamma}$. 

\begin{figure}[!t]
\vspace{-5pt}
\captionsetup[subfloat]{farskip=0pt,captionskip=-3pt} 
    \centering
  \subfloat[4 QAM]{%
       \includegraphics[width=0.95\linewidth]{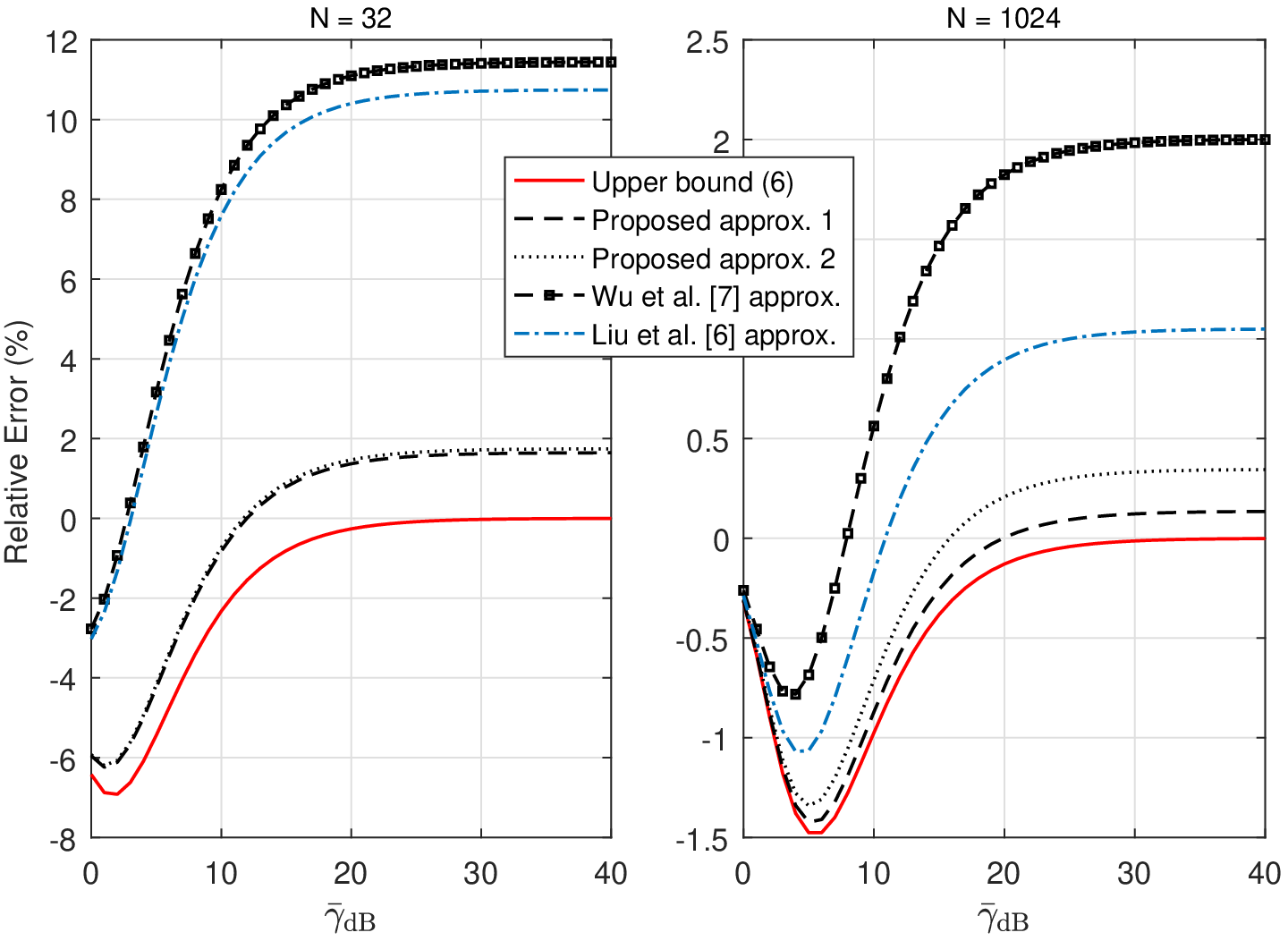}}
    \label{fig:4QAM}\\
  \subfloat[16 QAM]{%
        \includegraphics[width=0.95\linewidth]{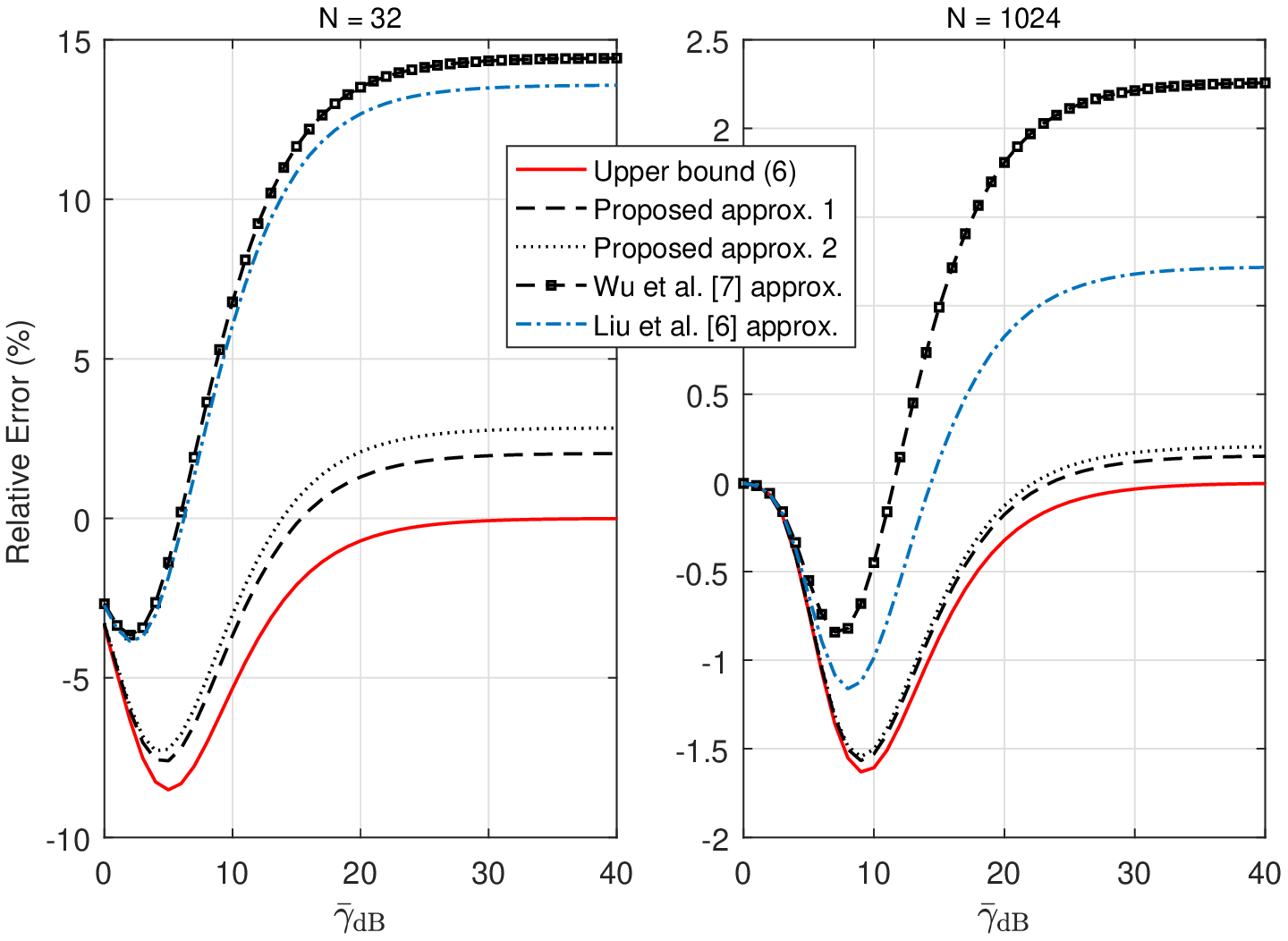}}
    \label{fig:16QAM}
  \caption{Relative error in average PER in Rayleigh fading. Approx. 1 uses $a_N$ and 
$b_N$ from \eqref{eq:AnBn_Qfunc} while Approx. 2 is based on \eqref{eq:AnBn_expo2}.}
  \label{fig:w0_perCompare} 
	\vspace{-13pt}
\end{figure}
We evaluate the average PER in \eqref{eq:UB_Naka} with the proposed 
$\omega_m$ approximation in \eqref{eq:proposedThreshold} with the 
original \eqref{eq:AnBn_Qfunc} and modified \eqref{eq:AnBn_expo2} parameters and validate it 
against the numerical evaluation of PER in
\eqref{eq:ref_PER}. The approximations to \eqref{eq:UB_Naka} for Rayliegh fading channel  
proposed in \cite{liu2012tput}\cite{wu2014energy} are also 
analyzed. Fig.~\ref{fig:w0_perCompare}(a) and Fig.~\ref{fig:w0_perCompare}(b) show the 
relative error (RE) in the proposed approximation and the reference studies 
for 4-QAM and 16-QAM in Rayleigh fading. Similar results (not 
shown here) are obtained for 64-QAM. 
It can be observed that the RE in average PER with 
$\omega_0$ approximations \eqref{eq:AnBn_Qfunc} and \eqref{eq:AnBn_expo2} 
are quite close to the upper bound \eqref{eq:UB_Ray}, which is evaluated numerically, for small to 
large packet lengths. In comprison, the RE of approximations in 
\cite{liu2012tput}\cite{wu2014energy} is small at low SNR, however it 
increases rapidly especially for small packet lengths. We also evaluated the 
average PER based on $\omega_m$ approximations in 
\eqref{eq:1} and \eqref{eq:2} and observed the similar accuracy.



\section{Energy Efficient Link Optimization}
\label{sec:EE}
\subsection{Energy Consumption Model}

We consider minimizing energy consumption of a wireless link between a 
transmitter and receiver pair separated by distance $d$. The energy 
consumption of the signal path at the transmitter and receiver is comprised of 
baseband processing blocks (e.g., (de)coding and (de)modulation) and 
radio-frequency (RF) chain that consists of a power amplifier (PA) and other 
electronic components such as analog-to-digital and digital-to-analog (AD/DA)
   converter, low-noise amplifier (LNA), filters, mixers and frequency 
synthesizers. However for an energy-constrained wireless system (e.g., WSN), the 
energy consumption of RF chain is orders of magnitude larger than that of 
baseband processing components. The power consumption of PA is considered to 
be proportional of the transmit power $P_t$ such that $P_{\mathrm{PA}} = \frac
 {\xi}{\eta}P_t$, where $\eta$ is the drain efficiency of the power amplifier 
(PA)  and $\xi$ is the peak-to-average-power-ratio (PAPR). The 
PAPR depends on the modulation scheme and the associated constellation 
size. 
If baseband power consumption is neglected and the power consumption of all 
the other components in RF chain excluding PA is denoted as $P_c$, a simple 
power consumption model is $P_{\mathrm{on}} = \frac{\xi}{\eta}P_t + P_c$. 
From \cite{cui2005energy}, this model leads to the total energy consumption to 
transmit and receive a symbol as
\begin{equation}
E_{\mathrm{sym}} = \frac{\xi}{\eta} E_t + \frac{P_c}{R_s}
\label{eq:E_symbol}
\end{equation}
where $E_t$ is the average transmission energy of a symbol and $R_s$ the 
physical layer symbol rate. For FSK, BPSK and QPSK modulations $\xi = 1$
, for OQPSK $\xi = 2.138$, and for a square MQAM modulation 
$\xi = 3 (\sqrt{M} - \frac{1}{\sqrt{M}} + 1)$~\cite{cui2005energy}.

Let $E_b = {E_r}/{\log_2M}$ be the average received energy per uncoded 
bit where $E_r$ is the average received energy per symbol and $M$ is the 
constellation size, then the average SNR, $\bar{\gamma}$, at the receiver is 
\begin{equation}
\bar{\gamma} = \frac{E_r}{N_0\log_2M}
\label{eq:snr}
\end{equation}

Assuming a $\kappa$th-power path-loss model, the transmission energy at 
distance $d$ from \eqref{eq:snr} is expressed as \cite{cui2005energy}
\begin{equation}
E_t \triangleq E_r G_d = \Big(\bar{\gamma}N_0 \log_2 M\Big) G_d
\label{eq:E_t}
\end{equation}
where $G_d \triangleq G_1 d^\kappa M_\ell$ is the pathloss gain with $G_1$,
the gain factor at unit distance, depends on the transmit and 
receive antenna gains and carrier frequency, and $M_l$ the link margin.

In packet based wireless systems, the information bits are encapsulated into 
packets each carrying $n_p$ payload and $n_h$ overhead bits. The number 
of symbols in a packet are $n_s = (n_h + n_p)/\log_2 M$. The average energy 
required to transmit and receive an information bit per packet transmission, 
from \eqref{eq:E_symbol} and \eqref{eq:E_t}, is
\begin{equation}
E_{\mathrm{0}} = \frac{n_s}{n_p}E_{\mathrm{sym}} = \frac{n_p + n_h}{n_p} A\bar{\gamma}  + B   
\label{eq:E_0}
\end{equation}
where $A = {\xi N_0 G_d}/{\eta}$ and $B = \frac{n_s}{n_p}\cdot \frac{P_c}{R_
s} = \frac{P_c}{R_b}$ with $R_b = W \log_2M$ the physical layer bit rate in bandwidth $W$.

The total energy consumption of a wireless link depends on the required 
retransmissions before a packet is decoded successfully at the receiver. The 
retransmission statistics are determined by the PER, 
$P_e\left(\bar{\gamma}\right)$, which is a function of $\bar{\gamma}$, 
channel fading, and other parameters as discussed earlier. The number of 
retransmissions $\tau$ is geometric random variable and over an uncorrelated 
channel between retransmissions the average number of retransmissions are $
\bar{\tau}  = {1}/{(1-\bar{P}_e(\bar{\gamma}))}$. Therefore, the total average 
energy for a successful transmission of a bit is $E = \bar{\tau}E_0$, which 
from \eqref{eq:E_0} is
\begin{equation}
E = \frac{1}{1-\bar{P}_e(\bar{\gamma})} \left(\frac{n_p + n_h}{n_p} A\bar{\gamma} + B\right)
\label{eq:E}
\end{equation}

In formulating \eqref{eq:E}, no limit on the number of retransmissions is 
assumed. However for a delay constrained system, a packet must be delivered 
within maximum number of retransmissions $\tau_r^{\max}$ and the 
packet error probability after $\tau_r^{\max}$ retransmissions must be less 
than a reliability target $\varepsilon$
\begin{equation}
\left[\bar{P}_e(\bar{\gamma})\right]^{\tau_r^{\max} + 1} \leq \varepsilon
\label{eq:PER_t}
\end{equation}
From \eqref{eq:PER_t}, the required PER $\varepsilon_{\mathrm{req}}$ to satisfy 
target $\varepsilon$ is
\begin{equation}
\bar{P}_e(\bar{\gamma}) \leq \varepsilon^{1/(\tau_r^{\max} + 1)} := 
\varepsilon_{\mathrm{req}}
\label{eq:PER_req}
\end{equation}

If \eqref{eq:PER_req} is satisfied, the average number of transmissions per packet is 
$\bar{\tau}_{\mathrm{trunc}}  = {1-[\bar{P}_e(\bar{\gamma})]^{\tau_r^{\max} + 1}}/{(1-\bar{P}_e(\bar{\gamma}))}$ and the total average energy is given by 
\begin{equation}
E_{\mathrm{trunc}} = \frac{1-[\bar{P}_e(\bar{\gamma})]^{\tau_r^{\max} + 1}}{1-
\bar{P}_e(\bar{\gamma})} \left(\frac{n_p + n_h}{n_p} A\bar{\gamma} + B\right)
\label{eq:E_trunc}
\end{equation}    

In next section, we consider minimizing energy consumption per information 
bit in \eqref{eq:E} while maintaining the PER constraint in 
\eqref{eq:PER_req}. 

\subsection{Link Optimization with Minimum Energy Consumption}

\subsubsection{Optimal Average SNR}
With $n_p$ fixed, finding the optimal average SNR represents a case where the 
sensors have to send a fixed size reports.  
The unconstrained energy minimization problem for optimal $\bar{\gamma}$ is 
modeled as
\begin{equation}
\begin{aligned}
& \underset{\bar{\gamma}}{\text{minimize}}
& & E(\bar{\gamma}) \\[-0.3em]
& \text{subject to}
& & \bar{\gamma} \in \left[0, \infty\right]
\end{aligned}
\label{eq:unConst_snr}
\end{equation}

The function $E$ is a product of two functions: $\bar{\tau}(\gamma)-$ the 
number of retransmissions with $\bar{\tau}^{'}(\gamma)\leq 0$, and 
$E_0(\gamma)-$ the average energy per transmission attempt such that 
$E_0^{'}(\gamma)\geq0$ where $\acute{x}$ denotes the first derivative. If both $\bar{\tau}(\gamma)$ and $E_0(\gamma)$ are 
convex, then $E$ is also convex \cite[Lemma 1]{wu2014energy} and the optimal 
$\bar{\gamma}$ can be obtained by solving $\frac{\partial E} {\partial \bar{\gamma}} = 0$ which yields a quadratic equation 
with a positive root as 
\begin{equation}
\bar{\gamma}^{*} = \frac{\omega_0}{2} + \sqrt{\omega_0 \left(\frac{\omega_0}{4
} + \frac{B}{A}\frac{n_p}{n_h+n_p}\right)}
\label{eq:opt_snr}
\end{equation}

Under the constraints on required PER and the transmit 
power, the minimization of energy in \eqref{eq:E} can be written as
\begin{equation}
\begin{aligned}
& \underset{\bar{\gamma}}{\text{minimize}}
& & E(\bar{\gamma}) \\[-0.3em]
& \text{subject to}
& & \bar{\gamma}_{\min}\leq\bar{\gamma} \leq \bar{\gamma}_{\max}
\end{aligned}
\label{eq:Const_snr}
\end{equation}
The minimum average SNR $\bar{\gamma}_{\min}$ requirement is set by the PER 
bound in \eqref{eq:PER_req}, which can be obtained from \eqref{eq:ubPER1}  
\begin{equation}
\bar{\gamma}_{\min} = - \frac{\gamma_e + \log\Big(\acute{c}_m\left(n_h+n_p\right)
\Big)}{\acute{k}_m \log\left(1-\varepsilon_{\mathrm{req}}\right)}
\label{eq:snr_min}
\end{equation}
Due to the hardware and regulatory constraints, the transmission power cannot 
exceed a limit $P_0$. The condition $P_{\mathrm{tx}} \leq P_0$ translates to 
$\bar{\gamma} \leq \bar{\gamma}_{\max}$ with $\bar{\gamma}_{\max}$ from \eqref{eq:E_t} is
\begin{equation}
\bar{\gamma}_{\max} = \frac{P_0}{WN_0G_d\log_2 M} 
\label{eq:snr_max}
\end{equation}

From \eqref{eq:snr_min} and \eqref{eq:snr_max}, the required SNR, denoted as 
$\bar{\gamma}^{*}_{\mathrm{req}}$, relates to the SNR for unconstrained case 
in \eqref{eq:unConst_snr} as 
\begin{align}
\bar{\gamma}_{\mathrm{req}}=
 \begin{dcases}
\bar{\gamma}_{\min}, & \bar{\gamma}^{*} < \bar{\gamma}_{\min}
\\
\bar{\gamma}_{\max}, & \bar{\gamma}^{*} > \bar{\gamma}_{\max}\\
\,\, \bar{\gamma}^{*}, &  \mathrm{otherwise}\\
 \end{dcases}
\label{eq:H_n_x}
\end{align} 
which holds for $\bar{\gamma}_{\min} < \bar{\gamma}_{
\max}$. If $\bar{\gamma}_{\min} > \bar{\gamma}_{\max}$, the reliability 
target cannot not be satisfied for a given modulation scheme. 

\subsubsection{Optimal Payload Size} 
The function $E$ in \eqref{eq:E} is also convex in payload size $n_p$ and its 
optimal value is
\begin{equation}
n_p^{*} = \!\!\frac{n_h \bar{\gamma}\left(\left(\acute{k}_m\! -1\right)\bar{\gamma}\! + \!\sqrt{
\acute{k}_m^2\bar{\gamma}^2\! + \! 2 \acute{k}_m \bar{\gamma}\! + \!\frac{4Bk_m}{A}\! + \!1}\right)}{2\Big(
\bar{\gamma} + \frac{B}{A}\Big)}
\label{eq:opt_L}
\end{equation}


The upper limit on the payload size ${n_{p,\max}}$ is set by the minimum SNR requirement $\bar{\gamma}_{\min}$ to satisify PER target. It is given by from \eqref{eq:PER_req}
\begin{equation}
{n_{p,\max}} = - n_h + \frac{10^{-(\gamma_e + \bar{\gamma}_{\min} \acute{k}_m \log\left(1-\varepsilon_{\mathrm{req}}\right))}}{\acute{c}_m}
\label{eq:max_L}
\end{equation}
where $\bar{\gamma}_{\min}$ is given in \eqref{eq:snr_min}.

\subsubsection{Joint Optimal $\bar{\gamma}, n_p, M, \tau_{r}^{\max}$} 

As the IoT devices will be used in diverse scenarios, it might be important in many to find the optimal SNR, payload size, modulation order and number of retransmissions for energy efficient communication. For example after deployment in harsh and inaccessible areas, the devices will optimize those parameters for the first time and then can continue with the optimal setting. The joint optimization problem can be written as 
\begin{equation}
\begin{aligned}
& \underset{\bar{\gamma}, n_p, M, \tau_{r}^{\max}}{\text{minimize}}
& & E\Big(\bar{\gamma}, n_p, M, \tau_{r}^{\max}\Big)
\end{aligned}
\end{equation}
where $M \in \{\mathrm{FSK, MPSK, MQAM}\}$ and $\tau_{r}^{\max} = i, \; i \geq 1$.

%
%

Note that these devices will support only few values of $M$ and a small value of 
$\tau_{r}^{\max}$ is feasible for minimum energy operation 
\cite{wu2014energy}. As a result, the exhaustive search over the combination of $M$ and $
\tau_{r}^{\max}$ will not be computationally demanding. For each combination of $M$ and $\tau_{r}^{\max}$, the joint optimum $\bar{\gamma}$ and $n_p$ can be found 
from \eqref{eq:opt_snr} and \eqref{eq:opt_L} either by solving system of two 
non-linear equations or by iteratively invoking these equations. In either case, we need to ensure that the reliability conditions in 
\eqref{eq:H_n_x} and \eqref{eq:max_L} are satisfied. 
However, the former method requires numerical evaluation that might be 
computationally infeasible for hardware-constrained devices. On the other 
hand by iteratively invoking \eqref{eq:opt_snr} and \eqref{eq:opt_L}, 
$\bar{ \gamma}$ and $n_p$ can efficiently converge to joint energy optimum 
values while satisfying the reliability conditions. It is straightforward to 
develop the proof of convergence of the iterative approach by following 
\cite[Corollary 3]{wu2014energy}. Note that by initializing $n_p$ and 
$\bar{\gamma}$ to any value, this approach converges within a few iterations 
to optimum values. A pseudocode of the proposed joint optimization is given in Algorithm~1.

\setlength{\textfloatsep}{4pt}
\begin{algorithm}[t]
\small
 \caption{Joint Optimization with Reliability Target}
 \begin{algorithmic}[1]
 \renewcommand{\algorithmicrequire}{\textbf{Input:}}
 \renewcommand{\algorithmicensure}{\textbf{Output:}}
 \REQUIRE $\varepsilon_{\mathrm{req}}, \tau_r^{\max}, \delta$
 \ENSURE  $\bar{\gamma}^{*}, n_p^*, \tau_r^*, M^*$
		\FOR {$M \in \left[\mathrm{FSK, MPSK, MQAM}\right]$}
			\FOR {$i = 1$ to $\tau_r^{\max}$}
				\STATE $n_p \gets 0$	
					\WHILE {$\Delta > \delta$}
						\STATE  $\bar{\gamma} \gets$ Evaluate (32), $\bar{\gamma}_{\min} \gets$ Evaluate (34), \\ $\bar{\gamma}_{\max} \gets$ Evaluate (35), $n_{p,{\max}} \gets$ Evaluate (38) 
						\IF {($\bar{\gamma}_{\min}> \bar{\gamma}_{\max}$)}
					       \STATE $\mathrm{break}$;
						\ELSE
					     	\STATE $ \bar{\gamma}_{\mathrm{req}} \gets $ Evaluate (36)  
					 	\ENDIF
								\STATE  $n_p \gets$ Evaluate (37) with $\bar{\gamma}$ =  $\bar{\gamma}_{\mathrm{req}}$
            \IF {$(n_p>n_{p,{\max}})$}
						    \STATE ($n_p \gets   n_{p,{\max}}$)
					  \ENDIF
								\STATE 	$E  \gets$ Evaluate (30) Print $E,{\gamma}, n_p,\tau_r, M$
								\STATE  $\Delta  \gets \bar{\gamma}_{\mathrm{req}} - \bar{\gamma}^{'}, \quad \bar{\gamma}^{'} = \bar{\gamma}_{\mathrm{req}}$
					\ENDWHILE
			\ENDFOR		
		\ENDFOR
 \RETURN $\bar{\gamma}, n_p, \tau_r, M$ yielding minimum $E$
 \end{algorithmic}
\vspace{-3pt} 
 \end{algorithm}

\subsection{Numerical Results}
\begin{figure}[t]
	\centering
		\includegraphics[width=1\linewidth]{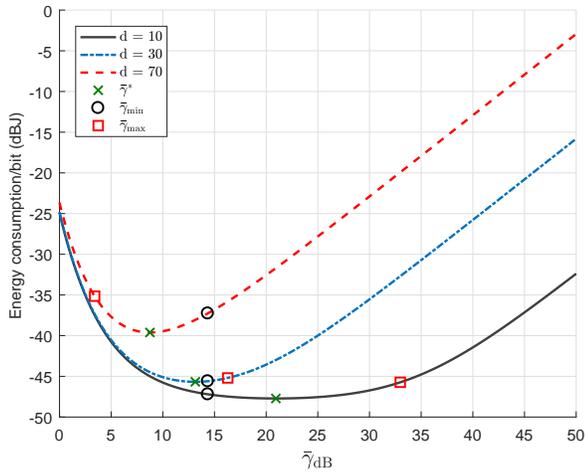}
		\vspace{-18pt}
	\caption{Optimal SNR vs required SNR for 4-QAM under reliability constraints of $
\varepsilon=0.001$, $\tau_r^{\max}=3$ and maximum transmit power of 
$P_0 = 10$mW.}
	\label{fig:opt_snr}
	\vspace{-14pt}
\end{figure}

The simulation parameters are taken from \cite{cui2005energy}: $N_0/2 = -174$
dBm/Hz, $\kappa = 3.5$, $G_1 = 30$dB, $M_{\ell}=40$dB, $W= 10$kHz, 
$P_c^{\mathrm{\{MQAM, MPSK\}}} = 310$mW, $P_c^{\mathrm{FSK}} = 265$mW, $\eta = 35\%$.
Other parameters are: $n_p = 48$ bits, $\varepsilon = 0.001$ (i.e., 99.9\% 
reliability), $P_0=10$mW. Fig.~\ref{fig:opt_snr} shows an example case of minimum SNR $\bar{
\gamma}_{\min}$ required at various distances. The unconstrained optimal SNR $\bar{\gamma}^*$ and 
maximum achievable SNR $\bar{\gamma}_{\max}$ are also depicted. At $d=10$, 
$\bar{\gamma}_{\min}$ is less than $\bar{\gamma}^*$, therefore $\bar{\gamma}^*
$ is energy optimal and is preferred over $\bar{\gamma}_{\min}$. While at $d = 30$, $\bar{\gamma}^*$ cannot satisfy the target and $\bar{\gamma}
_{\min}$, though not energy optimal, is selected. At $d=70$, the reliability 
target is not satisfied as $\bar{\gamma}_{\min}> \bar{\gamma}_{\max}$.  

\begin{figure}[t]
	\centering
		\includegraphics[width=1\linewidth]{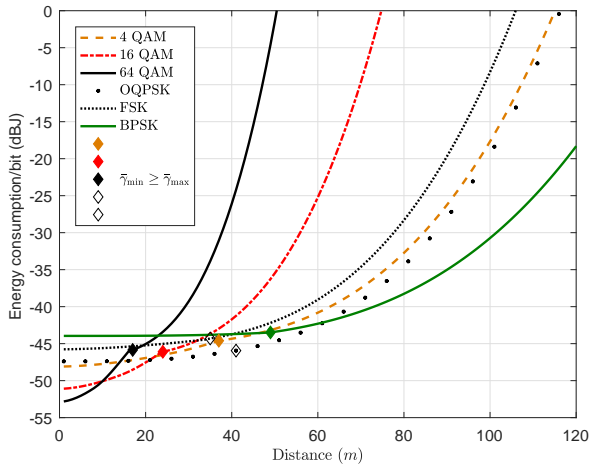}
	\vspace{-18pt}
	\caption{Energy consumption of the modulation schemes with optimal required SNR at each 
distance. The marked condition $\bar{\gamma}_{\min} \geq \bar{\gamma}_{\max}$ 
shows the distance beyond which the reliability constraints are not satisfied. Simulation parameters: $P_0 = 10$mW, $\varepsilon=0.001$, $\tau_r^{\max}=3$, $n_p = 984$, $n_h=40$.}
	\label{fig:snr_d}
\end{figure}

In Fig.~\ref{fig:snr_d}, energy consumption for selected modulation schemes 
with respect to distance when operated at optimal required SNR $\bar
{\gamma}_{\mathrm{req}}$ is shown. The condition at which $\bar{\gamma}_{\min}$ 
cannot be satisfied at a given transmit power constraint is also 
depicted. In addition, for $\bar{\gamma}_{\min} > \bar{\gamma}_{\max}$, we 
set ${\gamma}_{\mathrm{req}} = {\gamma}_{\mathrm{\max}}$ to depict the 
energy consumption under unlimited retransmissions. It is observed that 
there is an optimal modulation scheme at each 
distance that also satisfies the reliability target: high-order modulations 
at lower distance and low-order at higher distance as shown without 
reliability constraints in 
\cite{rosas2012modulation}. However for given transmit power limit, the 
distance at which the reliability target is satisfied decreases as the 
reliability requirement becomes tight. 

\begin{figure}
	\centering
		\includegraphics[width=1\linewidth]{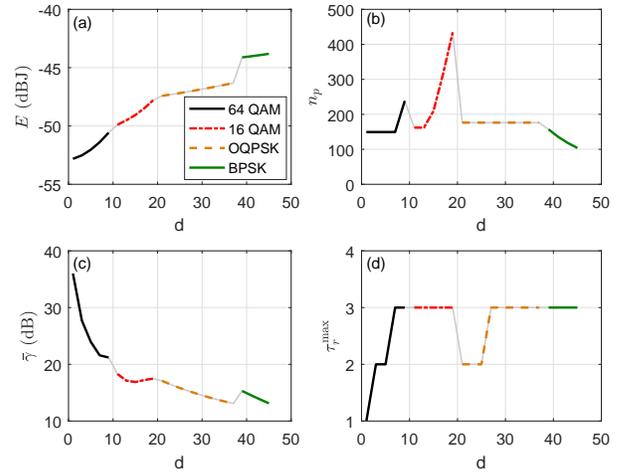}
		\vspace{-20pt}
	\caption{Optimal parameters at the output of the joint optimization 
algorithm: (a) energy consumption, (b) payload size (bits), (c) SNR, (d) number of 
retransmissions. The simulation parameters are the same as in Fig.~\ref{fig:snr_d}.}
	\label{fig:algo}
\end{figure}

In Fig.~\ref{fig:algo}, one can grasp the big picture of how the parameters $\bar{\gamma}, n_p, M$ and $\tau_{r}^{\max}$ vary with distance. At very short distance high $M$ and lower $n_p$ are energy efficient. The reason behind lower $n_p$ can be explained with the smaller value of $\tau_r^{\max}$. As distance increases optimal $M$ becomes smaller. The payload size $n_p$ keeps increasing at around $7-8$m and $13-19$m region with the increase in distance keeping the $\bar{\gamma}$ almost constant, i.e., increasing transmit power with increasing packet size is optimal until next smaller $M$ becomes energy optimal. At long distances lower $M$ and $n_p$ are energy optimal.
  
\vspace{-8pt}
\section{Conclusions} 
\label{sec:Conclusions}
In this paper, we studied the cross-layer link optimization while ensuring 
energy efficiency and reliability constraints. For cross-layer analysis, we 
first presented a simple approximation to average PER in block fading 
channels. The proposed PER approximation is in the form elementary functions, 
and maintains an explicit connection between the physical/link layer 
parameters and the packet error rate. The numerical analysis confirms the 
tightness of the approximation as compared to earlier studies. Later, we 
exploited the proposed PER approximation in the energy consumption model to
find energy optimal yet reliability and hardware compliant conditions 
for unconstrained optimal SNR and payload size. These conditions are shown to 
be useful to: i) find optimal SNR for a system with fixed modulation scheme and 
payload size, ii) develop an holistic algorithm to jointly optimize the 
physical and link layer parameters.   

\vspace{-2pt}
\bibliographystyle{IEEEtran}
\bibliography{GC17}

\end{document}